\documentclass[%
 reprint,
 amsmath,amssymb,
 aps,
]{revtex4-1}

\usepackage{graphicx}
\usepackage{xcolor}
\usepackage{braket}
\usepackage{dcolumn}
\usepackage{bm}
\usepackage[utf8]{inputenc}
\usepackage{pdfcomment}
\usepackage{hyperref}

\begin{document}

\preprint{APS/123-QED}

\title{Ultrastrongly dissipative quantum Rabi model}

\author{David Zueco}
 \affiliation{Instituto de Ciencia de Materiales de Aragon and Departamento de Física de la Materia Condensada, CSIC-Universidad de Zaragoza, E-50009 Zaragoza, Spain}
\affiliation{Fundación ARAID, Paseo María Agustín 36, E-50004 Zaragoza, Spain}
\author{Juanjo García-Ripoll}%
\affiliation{Instituto de Física Fundamental, IFF-CSIC, Calle Serrano
 113b, Madrid E-28006, Spain}

\date{\today}

\begin{abstract}

  We discuss the equilibrium and out of equilibrium dynamics of cavity QED in presence of dissipation  beyond the standard perturbative treatment of losses.  Using the dynamical polaron \emph{ansatz} and Matrix Product State simulations, we discuss the case where both  light-matter $g$-coupling  and system-bath interaction  are in the ultrastrong coupling regime.
  We provide a critical $g$  for the onset of Rabi oscillations. 
Besides, we demonstrate that the qubit is \emph{dressed} by the cavity and dissipation. That such  dressing governs the dynamics and, thus, it can be measured. Finally, we sketch an implementation for our theoretical ideas within circuit QED technology.
\end{abstract}

\pacs{Valid PACS appear here}
\maketitle



\section{\label{sec:intro}Introduction}

\begin{figure}[b!]
\includegraphics[width=0.9\linewidth]{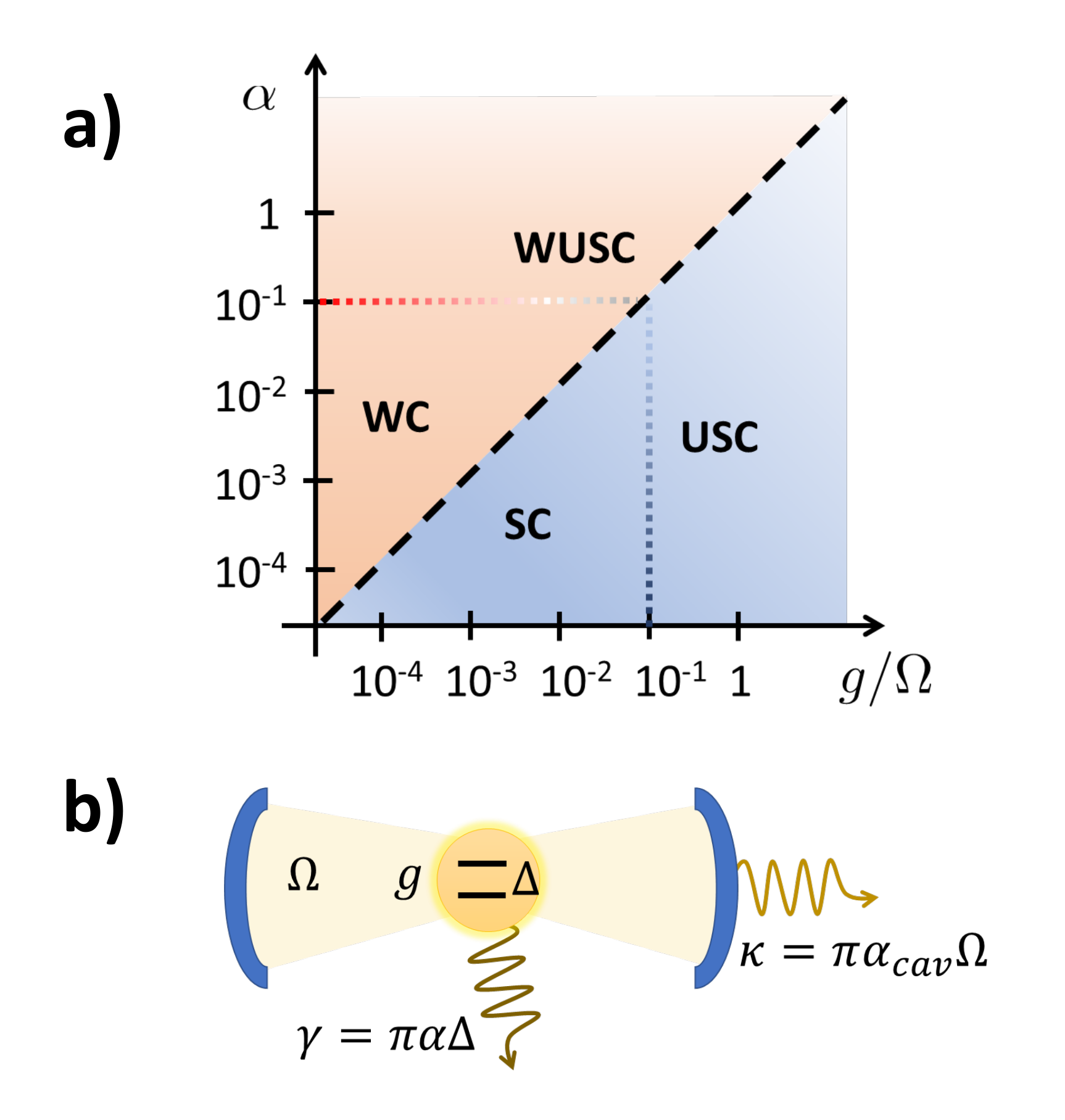}
\caption{Cavity QED phase diagram and setup sketch.
a) Blue region marks where Rabi oscillations occur. We distinguish between strong (SC) and ultrastrong (USC) coupling regimes. In the red region, losses are big enough and the TLS decays in an overdamped way. Here, we 
 distinguish between weak (WC) and weak ultrastrong coupling regime
 (WUSC). In the figure $\alpha$ characterizes the TLS losses
 [Cf. Eq. \eqref{gamma}]. In b) we draw a cavity QED skectch with the main parameters indicated.}
\label{fig:pd}
\end{figure}

More than 80 years ago, Rabi studied the interaction of a two level system (TLS) with a classical electromagnetic field \cite{Rabi1936}. Jaynes and Cummings (JC) quantized this theory\ \cite{Jaynes1963}, focusing on the case of a single mode. This model is  the \emph{quantum Rabi model} $(\hbar=1)$
\begin{equation}
\label{qR}
H_{\rm qR} = \frac{\Delta}{2} \sigma_z + \Omega a^\dagger a  + g \sigma_x  (a^\dagger + a)  \,.
\end{equation}
In this Hamiltonian,  $\Delta$ and $\Omega$ are the \emph{bare} TLS and cavity frequencies, while $g$ denotes the light-matter interaction strength, see Fig. \ref{fig:pd}b. Considering the dissipation of the two-level system $\gamma$ and of the cavity $\kappa$, we obtain several light-matter regimes [Cf. Fig.\ \ref{fig:pd}a]. When coupling outweighs dissipation $g\gg \{\gamma, \kappa \}$, the qubit and the cavity field exchange excitations in a coherent way. Here (blue zone in Fig. \ref{fig:pd}a) we distinguish two regimes. If $g/\Delta \lesssim 0.1$, we are in the \emph{strong coupling (SC) } regime and light-matter interaction can be simplified into $g (\sigma^+ a + {\rm h.c.})$ using the Rotating Wave Approximation (RWA). However, if $g/\Delta \gtrsim 0.1$, the RWA fails and the full interaction, \emph{i.e.} the counter rotating terms ---$g (\sigma^+ a^\dagger + \rm{h.c.})$--- are needed. This is the \emph{ultrastrong coupling regime (USC)} \cite{Forn2018}. By analogy, when dissipation dominates, we distinguish between the \emph{weak (W)} and \emph{weak ultrastrong coupling (WUSC)}, depending on whether we can apply the Markovian approximation or not to describe dissipation (red area in Fig. \ref{fig:pd}a) .

The goal of this work is to derive a mathematical treatment of the cavity-QED model that provides quantitatively or qualitatively accurate solutions in all coupling and dissipation regimes --WC, SC, WUSC and USC--. There are many ways to solve the cavity-QED model that apply to subsets of these regimes. In absence of dissipation, equation \eqref{qR} admits an analytical solution \cite{Braak2011} and can be solved in the computer for any $g$-value. If losses are taken into account, they are typically discussed using Markovian master equations\ \cite{Breuer2007,Rivas2012}. They are perturbative in the system-bath interaction \cite{Ashhab2010, Beaudoin2011}. Going beyond this perturbative treatment is tricky \cite{Weiss2008}. Renormalization, path integral expansions or numerical techniques are required \cite{deVega2017,LeHur2017}. In contrast to this zoo of solutions, we will offer a unique method of broad utility with the only restriction that qubit dissipation remains below the quantum phase transition into the strongly correlated regime. The resulting method will be useful in studying all the quantum technologies that are developed around the JC model ---single photon emitters, quantum computers, spin squeezing\ \cite{Buluta2011}--, as well as experiments that exploit the huge dipole moments of superconducting qubits in the \emph{ultrastrong coupling} regime \cite{Niemczyk2010,Forn-Diaz2010,Yoshihara2016}.

Our method builds on the polaron Hamiltonian \cite{Bera2014, Camacho2016} to develop an effective model that can be analytically or numerically solved. Similar to the Ohmic spin-boson theory \cite{Leggett1987,Peropadre2013,Forn-Diaz2017}, we predict non-Markovian renormalization of the qubit splitting $\Delta$ due to the coupling with the bath, either directly $\gamma$ or via the cavity. We can also solve the qubit-cavity dynamics, from overdamped decay in the limit of WC or WUSC dissipation, to coherent scenarios that extend well inside the USC both in losses and light-matter coupling, see Fig. \ref{fig:pd}.

The outline of this work is as follows. In the next section \ref{sec:model}, we summarize the model, the polaron transformation and the different ways of solving the equilibrium and out of equilibrium dynamics. In section \ref{sec:results} we announce our results. We discuss the ground state properties of the model, the onset of Rabi oscillations and the noise spectrum. Finally, we give some conclussions and a possible implementation in \ref{sec:conclusions}. Several technical details are sent to the appendices.

\section{Theoretical methods}\label{sec:model}

\subsection{Combined dissipation channels}

We consider a qubit and a cavity interacting with each other and coupled to independent baths. The model in the system-bath formalism\ \cite{Caldeira1983, Hanggi2005} is [cf. Eq, \eqref{qR}],
\begin{align}
 \label{HT}
H & = H_{\rm qR} + \sum_{i=1,2} \sum_{k}^N \omega_{k,i} b_{k,i}^\dagger b_{k,i}
 \\ \nonumber
 &+
\sigma_x
\sum_k^N c_{k,1} 
X_{k,1}
+
(a + a^\dagger) \sum_k^N c_{k,2} 
 X_{k,2}
 \\ \nonumber
 &+ (a+a)^2 \sum_k \frac{|c_k|^2}{2 \omega_k}
 \;.
\end{align}
The qubit and cavity baths have independent modes $\omega_{k, i}$ ($i=1,2$), with bosonic quadratures $X_{k, i} \equiv b_{k,i}^\dagger + b_{k,i}$. Both noise channels can be described using spectral density functions,
$ J_i (\omega) = 2 \pi \sum_k c_{k,i}^2 \delta (\omega -\omega_{k,i})$.
In this paper we are considering Ohmic noise spectrum for both the cavity and the spin, {\emph i.e.} $J_i (\omega) \sim \omega$. In the Markovian limit, the dissipation strength is determined by the spontaneous emission rates of the qubit ($\gamma$) and the cavity $\kappa$ because [Cf. Fig. \ref{fig:pd}b]:
\begin{subequations}
 \begin{align}
 \label{gamma}
 \gamma & = J_1 (\Delta) = \pi \alpha \Delta \; ,
 \\
 \label{kappa}
   \kappa & = J_2 (\Omega) = \pi \alpha_{\rm cav} \Omega \; .
\end{align}
\end{subequations}
with $\alpha$ ($\alpha_{\rm cav}$) dimensionless parameters characterizing the dissipation strenght for the TLS (cavity).

The last term in equation\ \eqref{HT} deserves some discussion. This regularization of the bosonic modes arises from a cavity-bath coupling of the form $\sim (a+a^\dagger - \Phi)^2,$ where $a+a^\dagger$ is the cavity quadrature and $\Phi$ is the electromagnetic field injected by the bath . This type of coupling ---which is very natural in superconducting circuits--- ensures that the total energy is bounded from below and leads to the quadratic correction of the bosonic modes. Importantly, the quadratic correction ensures that the resonance of the cavity stays at $\Omega$, irrespective of the dissipation strenght $\alpha_{\rm cav}$. Note also that we do not find a similar term in the qubit-bath coupling because of saturation: $(\sigma^x)^2=1$. Further parameter renormalization is associated to quantum many body effects between the bath and the cavity QED system \cite{Ingold2002, Weiss2008}.

The cavity mode in \eqref{qR} can be diagonalized together with its environment. In doing so, Hamiltonian \eqref{HT} is rewritten as a spin-boson model\ \cite{Leggett1987} for a two level system coupled to two baths, one of which contains the cavity mode,
\begin{align}
\label{H-sb}
H = \frac{\Delta}{2} \sigma_z + \sigma_x \sum_{k'}^{2N+1} c_{k'} (b_{k'}^\dagger + b_{k'}) + 
 \sum_{k'}^{2N+1} \omega_{k'} b_{k'}^\dagger b_{k'}\;.
\end{align}
By joining the cavity modes and the qubit bath, we arrive at the total spectral density $J (\omega) = 2 \pi \sum_{k^\prime} c_{k^\prime}^2 \delta (\omega - \omega_{k^\prime})$:
\begin{equation}
\label{Jsb}
J(\omega) = 
\pi \alpha \omega + 
\frac{ 4 g^2 \pi \alpha_{\rm cav} \Omega^2
 \omega }{(\Omega^2-\omega^2)^2 + ( \pi \alpha_{\rm cav} \Omega \omega)^2} 
 \; .
\end{equation}
The second term characterizes the bath containing the cavity mode and is peaked around the cavity frequency $\Omega$ \cite{Garg1985}. The first term accounts for the intrinsic qubit Ohmic environment. We will correct this spectral function by introducing a  hard cut-off,  $\omega_c$ \citep{Peropadre2013}.

Finally, note that models, \eqref{HT} and \eqref{H-sb} are completely equivalent. When writing the model as \eqref{H-sb}, Rabi oscillations can be understood as non-Markovian decaying oscillations comming from the peaked spectral density. Details on the cavity-bath diagonalization and the effective spectral density are given in App. \ref {app:peaked}.

\subsection{Effective RWA models}

It has been recently shown that the low energy spectrum of a spin-boson model\ \eqref{Jsb} can be very well approximated by an effective, excitation number conserving Hamiltonian derived from a polaron transformation \cite{Bera2014, Camacho2016}. The basic idea is to construct a unitary transformation that disentangles the TLS from the bath
\begin{equation}
\label{Up}
U_p = e^{\sigma_x \sum (f_k b_k^\dagger - f_k^* b_k)} \; ,
\end{equation}
and choosing the displacements $f_k$ with the Silbey-Harris prescription that the ground state of $H_p=U_p^\dagger H U_p$ be as close as possible to $\ket{0}\otimes\ket{\mathbf{0}},$ the ground state of the uncoupled TLS $\ket{0}$ and of the bath $\ket{\mathbf{0}}$. Minimization yields the self-consistent relation
\begin{equation}
\label{fkDr}
f_k = \frac{-c_k/2}{\Delta_r + \omega_k} \; \; {\rm with} \; \; 
\Delta_r = \Delta \, e^{- 2 \sum_k f_k^2}
\end{equation}
and the effective Hamiltonian $H_p$ is well approximated, within the single-excitation sector, by \cite{shi2018}
\begin{align}
\label{Hp}
H_{p1} &\cong \frac{\Delta_r}{2} \sigma_z + \sum_k^N \omega_k b_k ^\dagger b_k
\\ \nonumber
& + 2 \Delta_r \big ( \sigma^+ \sum_k^N f_k b_k + {\rm h.c.} \big ) -
    2 \Delta_r \sigma_z \sum_{k, p}^N f_k f_p b_k^\dagger b_p
\; .    
\end{align}
This model has two important features. First, our TLS appears with the renormalized frequency $\Delta_r$, that determines the dynamics of the qubit in the bath [cf. Sect.\ \ref{sec:Dynamics}]. Second, the model develops a conserved quantity $[H_p, \sigma_z + \sum_k b_k^\dagger b_k]=0$ and becomes tractable with the same techniques as RWA models.

\subsection{Estimating $\Delta_r$}
\label{sect:Dr}

The qubit renormalized frequency, $\Delta_r$ [Cf. Eqs. \eqref{fkDr} and \eqref{Hp}] admits analytical solutions in the continuum limit, where we can write
\begin{equation}
 \label{Drcont}
\Delta_r = \Delta {\rm e}^{- 1/2\int_0^{\omega_c} J(\omega)/(\omega +
 \Delta_r)^2} \; .
\end{equation}
using the  UV cutoff $\omega_c$. The resulting expression is not tractable, and requires the numerical solution of a transcendental equation for $\Delta_r$ in \eqref{Drcont}. One extreme limit of this equation appears when the qubit decouples from the cavity ($g=0$). In that case Eq. \eqref{Drcont} can be solved, resulting in the Ohmic spin boson model. Then, $\Delta_r = \Delta(\Delta/\omega_c)^{\alpha/(1 -\alpha)} $, and the localization-delocalization transition at $\alpha=1$ \cite{Leggett1987}. When we depart from this limit $g\neq0$, the second summand in \eqref{Jsb}, decrease the onset of the localization transition to lower values of $\alpha$.

The predictions of the polaron ansatz can be compared with the adiabatic renormalization (ARG) \cite{Leggett1987} in the continuum limit. Applied to the spin-boson model, the ARG predicts a qubit frequency at
\begin{equation}
 \label{Drad}
 \Delta_r^{\rm ad}= \Delta {\rm e}^{- 1/2\int_{\Delta_r}^{\omega_c} J(\omega)/\omega ^2} \;. 
\end{equation}
Comparing \eqref{Drcont} and \eqref{Drad} we see a difference in the lower limit of the integral. The reason is that the RG flow stops at $\omega_c \sim \Delta_r$. If $g=0$, both methods yield the same result except corrections of the order of $\mathcal {O} (\Delta_r /\omega_c)$. Therefore, they predict the same $\Delta_r$. However, if $g\neq0$ ARG is different from polaron (and becomes less accurate). Our interpretation is that the cavity $\Omega$ behaves as an \emph{effective} cuttoff and for $\Delta \sim \Omega$ the ARG should fail. More generally, we also expect the ARG to fail whenever cavity losses and/or light-matter coupling dominate over the TLS intrinsic dissipation.

\subsection{Modified Wigner-Weisskopf}
\label{sec:wigner-weisskopf}

We have just discussed that \eqref{Hp} conserves the number of excitations. We can solve the dynamics within a single-excitation subspace \textit{{\`a} la} Wigner-Weisskop. There, the dynamics is fully determined by the wavefunction:
\begin{equation}
\label{dynP}
\ket{\Psi (t)} = U_p[f_k]^\dagger (\psi \sigma^+ + \sum_k\psi_k b_k^\dagger ) \ket{0,\mathbf{0}}
\; .
\end{equation}
Using \eqref{Hp} and \eqref{dynP}, the coefficients $\{\psi, \psi_k \}$ satisfy the set of coupled linear equations:
\begin{subequations}
\begin{align}
\label{WWPa}
\dot \psi &= - i 2 \Delta \sum \psi_k f_k
\\
\label{WWPb}
\dot \psi_k &= -i (\omega_k - \Delta) \psi_k - i 2 \Delta f_k \Big ( \psi + \sum_{k^\prime} f_{k^\prime} \psi_{k^\prime} \Big )
\; .
\end{align}
\end{subequations}
From these coefficients we may derive, for instance, the excitation probability of the two-level system
\begin{align}
\label{Pet}
 P_e = \frac{\langle \psi (t) | \sigma_z |
 \, \psi (t) \rangle + 1}{2}
\; .
\end{align}
In the regime $g/\Delta, \alpha, \alpha_{\rm cav} \ll 1$, we can solve for the
qubit amplitude $\psi$ applying the Markov approximation on the
qubit losses and replacing the second summand in \eqref{Jsb} with a
Lorentzian centered on the cavity resonance $\Omega$. Then an
analytical solution is possible as it is fully developed in our App. \ref{app:se}.
In the ultrastrong both for losses and light-matter coupling,, analytical advances are possible in the calculation of the qubit noise spectrum $S(\omega)$ as we explain in Sect. \ref{sec:Sw}. 
However, for the time evolution, in general, an analytical solution is no
longer possible. Then, we approximate the environment
using a finite number of modes, $N$, as explained in App.
\ref{app:discrete}. In doing so, we can solve the set of $\mathcal{O}(N)$ ordinary differential \eqref{WWPa} and \eqref{WWPb} numerically, {\emph e.g.} using Lanczos, Runge-Kutta or any other available method.

\subsection{Matrix Product States}
\label{sect:MPS}

In order to confirm the predictions of $H_{p}$, we also run numerical simulations on unapproximated model $H_P=U_p^\dagger H U_p$ using Matrix Product State ansatz. While working with $H_p$ significantly decreases the amount of entanglement in the MPS simulation, we introduce another optimization and express the $H_p$ as a tight-binding model\ \cite{chin2010}. The simulated model reads
\begin{align}
 H_p &= \frac{\Delta_r}{2}\sigma_z e^{-\theta c^\dagger_0}e^{\theta c_0} + 2\Delta_r \theta \sigma_x (c_0+c_0^\dagger) +\\
  &+ \sum_{i} (\beta_i c_{i+1}^\dagger c_i + \beta_i c_i^\dagger c_{i+1} + \alpha_i c_i^\dagger c_i),\nonumber
\end{align}
where $\theta^2 = \sum_k f_k^2.$
The new collective modes $c_i$ are constructed from the original ones $b_i$ using a Lanczos recursion\ \cite{chin2010} that also produces the real numbers $\alpha_i$ and $\beta_i$. This new model has the advantage that it can be simulated using both Arnoldi and Trotter-type MPS methods\ \cite{garcia2006}.

\section{Cavity-QED beyond Markovian regime}
\label{sec:results}

\subsection{Ground state}
\label{sec:Statics}

We will now show how to apply the previous formalism to study the static and dynamic properties of the cavity-QED setup in all regimes. We begin with the nature and properties of the ground state.

The construction of the polaron Hamiltonian provides a zeroth-order approximation to the ground state, $U_p\ket{0,\mathbf{0}},$ which predicts that the qubit has some probability to be excited. This is consistent with earlier findings in lossless cavities\ \cite{Irish2007, Ashhab2010} and in the spin-boson model without cavity\ \cite{Leggett1987}, but our new treatment allows us to interpolate between both limits. The equilibrium $z$-magnetization is proportional to the qubit renormalized frequency,
\begin{equation}
\label{sz}
\langle \sigma_z \rangle_{\rm eq}= \langle 0, {\bf 0} | U_p^\dagger \sigma_z U_p | 0, {\bf 0} \rangle =-\frac{\Delta_r}{\Delta}
\;,
\end{equation}
which as we saw before, can be computed from the displacement $f_k$ \eqref{fkDr}.

\begin{figure}[b!]
\includegraphics[width=1.\linewidth]{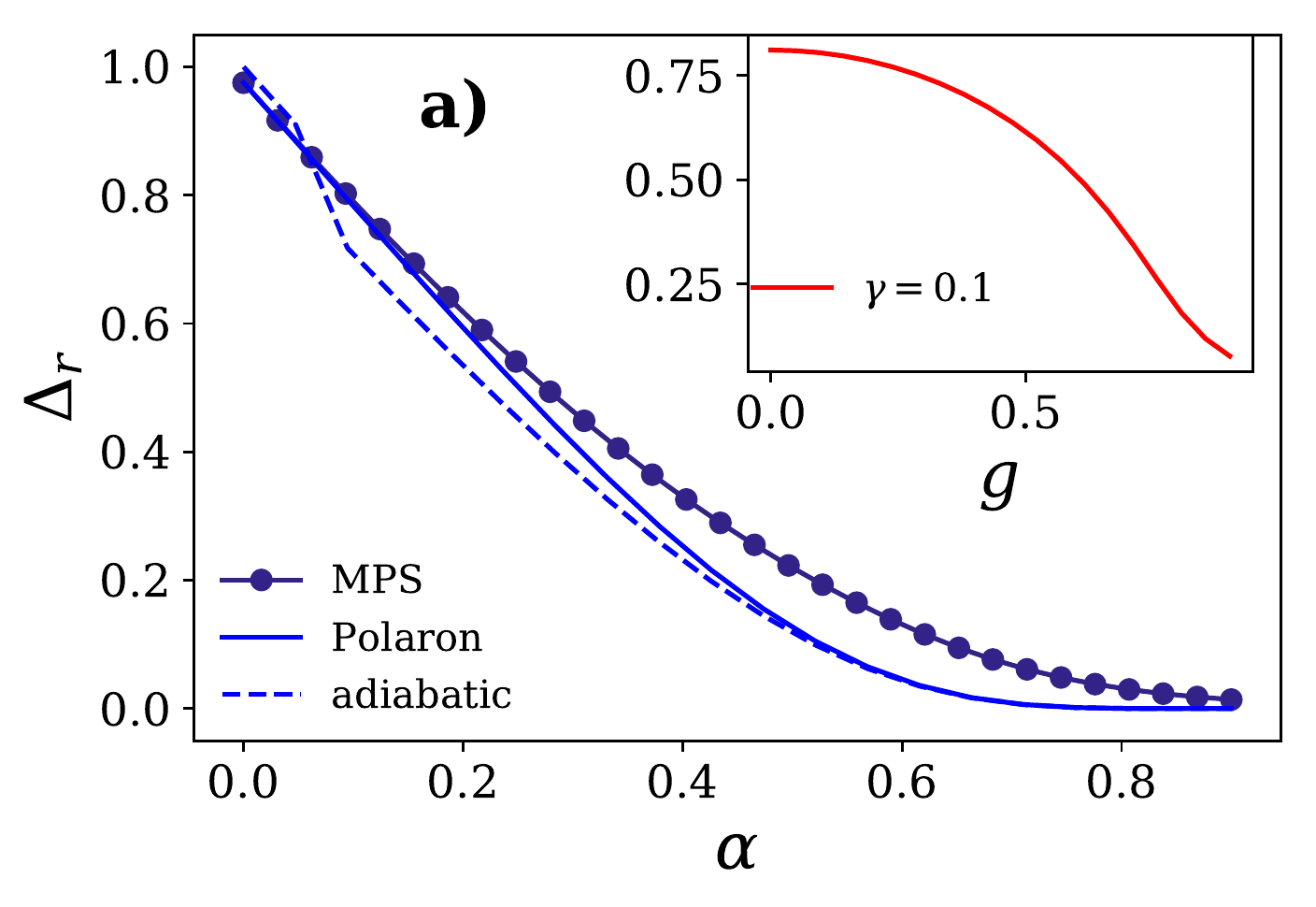}
\includegraphics[width=1.\linewidth]{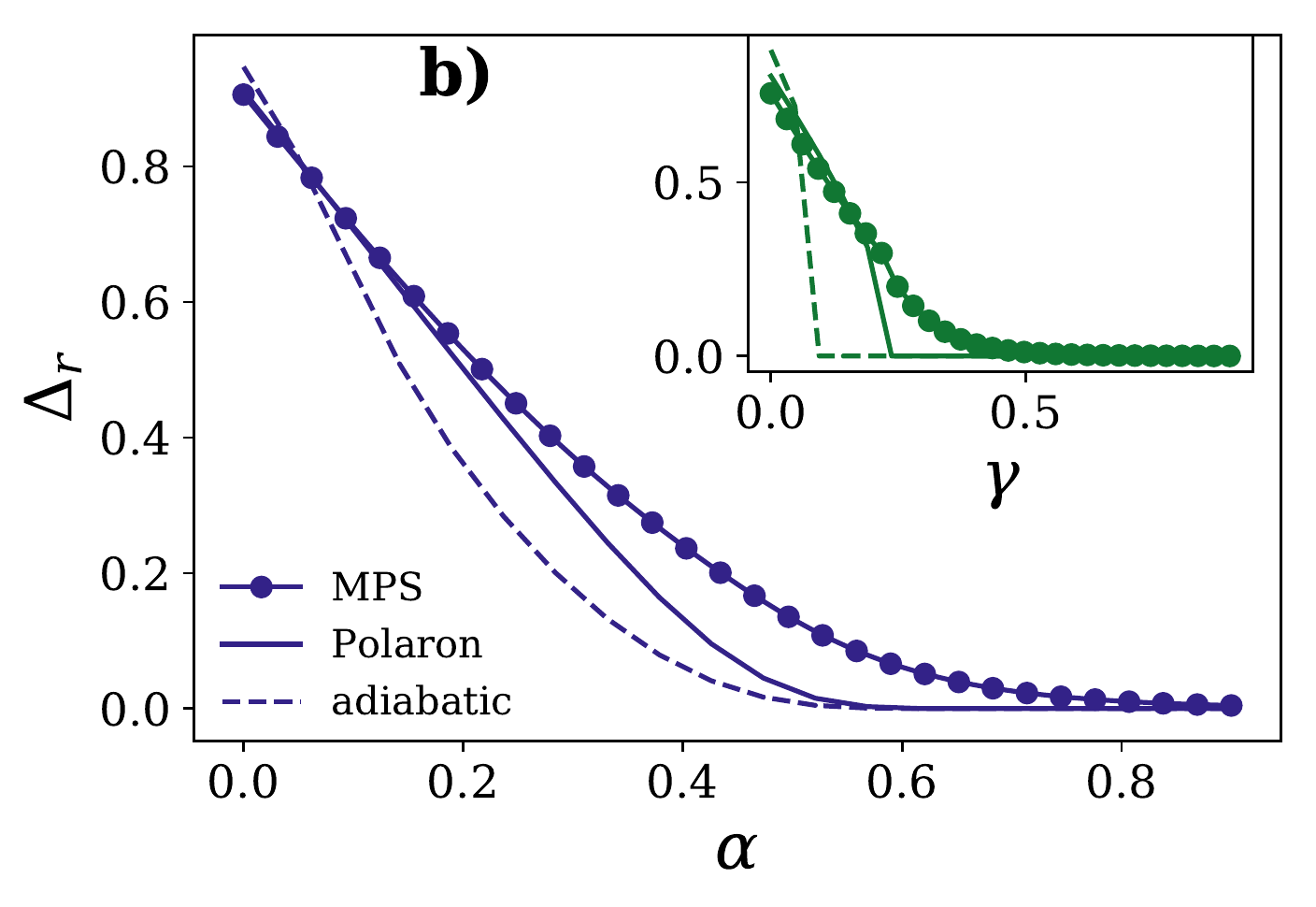}
\caption{Frequency renormalization ($\Delta_r$). In a) we compare the prediction of our three theories: polaron, adiabatic RG and MPS numerical simulation as a function of $\alpha$. We set $\kappa= \pi \alpha_{\rm cav}= 0.01 \pi $ and $g=0.2$. In the inset we show the dependence on $g$ strenght for $\alpha= 0.1$ and the same $\kappa$. In b) we do the same but setting $\kappa= \pi 0.8$ and $g=0.4$ (main panel) and $\kappa = \pi 1.5$ and $g=0.6$ (inset). In all the figures the cavity and qubit bare parameters are $\Omega = \Delta = 1$.}
\label{fig:Dr}
\end{figure}

Let us now compare the estimates of $\braket{\sigma_z}$ and $\Delta_r$, from the adiabatic renormalization group, \eqref{Drad}, the polaron method, Eq. \eqref{Drcont}, and an exact solution of $H_p$ with MPS discussed in Sects. \ref{sect:Dr} and \ref{sect:MPS}. Fig.\ \ref{fig:Dr} summarizes the ground state properties for different values of the dissipation and coupling strength. Those simulations have been performed with $N=256$ modes for the cavity bath, and a similar amount for the qubit bath, ensuring numerical convergence to a quasi-continuum limit. In panel \ref{fig:Dr}a we plot the qubit renormalization $\Delta_r$ as a function of the TLS dissipation [$\alpha$, Cf. Eq. \eqref{gamma}], for bare parameters $\Omega=\Delta=1$, an USC coupling strength $g=0.2$ and a low cavity spontaneous emission $\kappa= \pi \alpha_{\rm cav} \Omega = \pi 0.01$. We compare three methods: polaron, ARG and MPS simulations, as explained in Sect.\ \ref{sec:model}. The dependence of $\Delta_r$ on $\alpha$ resembles the pure Ohmic spin-boson model. As we have anticipated and explained in Sect. \eqref{sect:Dr} ARG is not accurate for small TLS intrinsic noise strenght $\alpha$. The inset of Fig. \ref{fig:Dr} also shows that the qubit-cavity coupling lowers even further the qubit frequency $\Delta_r$, due to the friction induced by the additional bosonic modes from the cavity and its bath..

Figure \ref{fig:Dr}b probes the ground state for stronger cavity dissipation, entering the WUSC regime. As seen on the main panel, for relatively high cavity losses ($\kappa= \pi \alpha_{\rm cav} \Omega=\pi 0.8$) the trend of $\Delta_r$ is qualitatively similar to the uncoupled case $g=0$. However, if we increase $\kappa$ (and $g$) enough  both the polaron and ARG models predict a sharp transition, leading to localized solutions $\Delta_r=0$. Since the MPS are numerically  exact simulations and do not exhibit such a transition, we conclude that this is an artifact of the polaron method that constraints its applicability to large values of dissipation and light-matter coupling, $g\geq 0.6$ and $\kappa \geq 1$. The remaining of this work will stay well within this regime, in which simulations verify well against MPS.

\subsection{Non perturbative Rabi oscillations}
\label{sec:Dynamics}

\begin{figure*}
 \includegraphics[height=.28\linewidth]{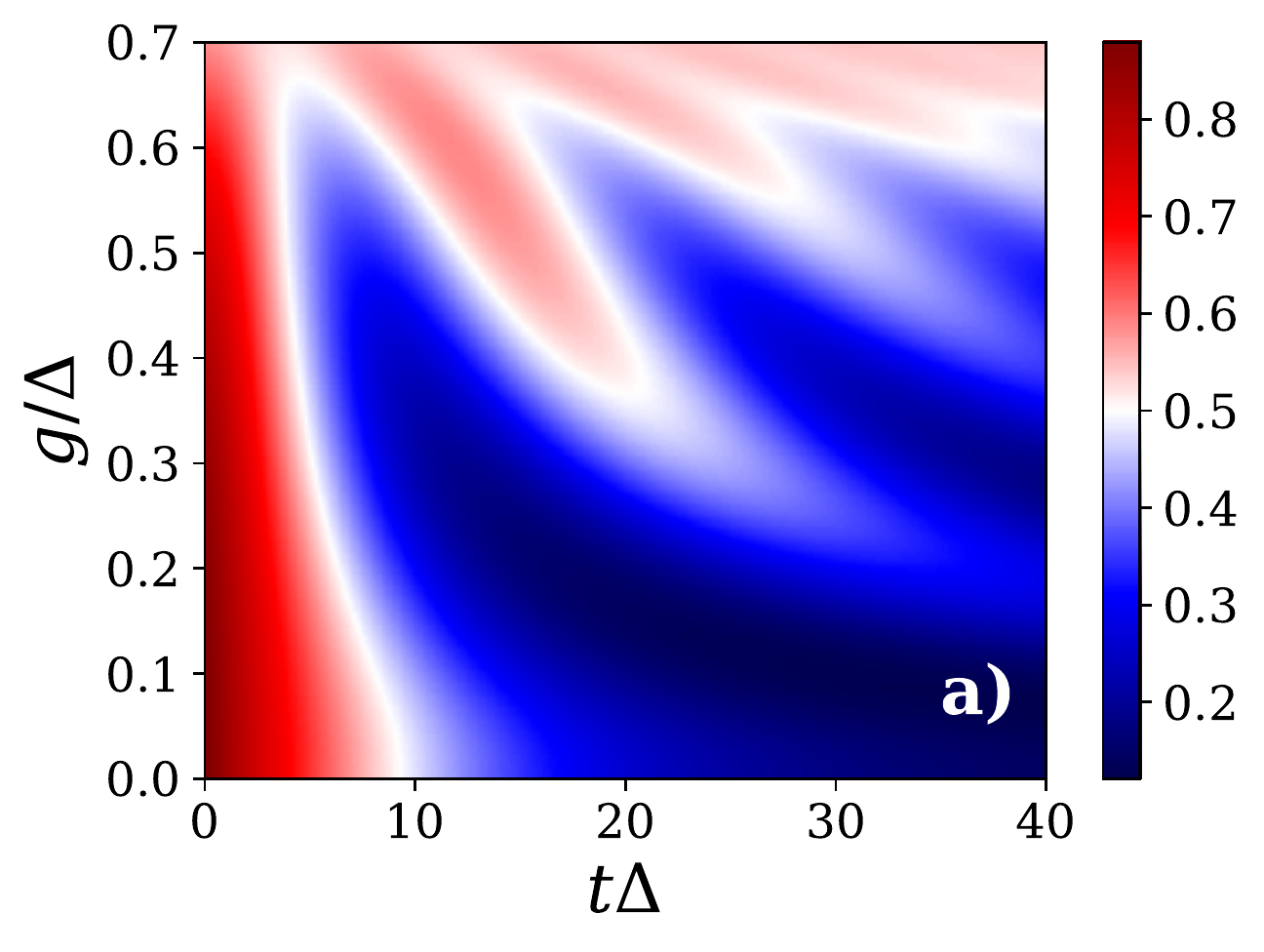}\includegraphics[height=.28\linewidth]{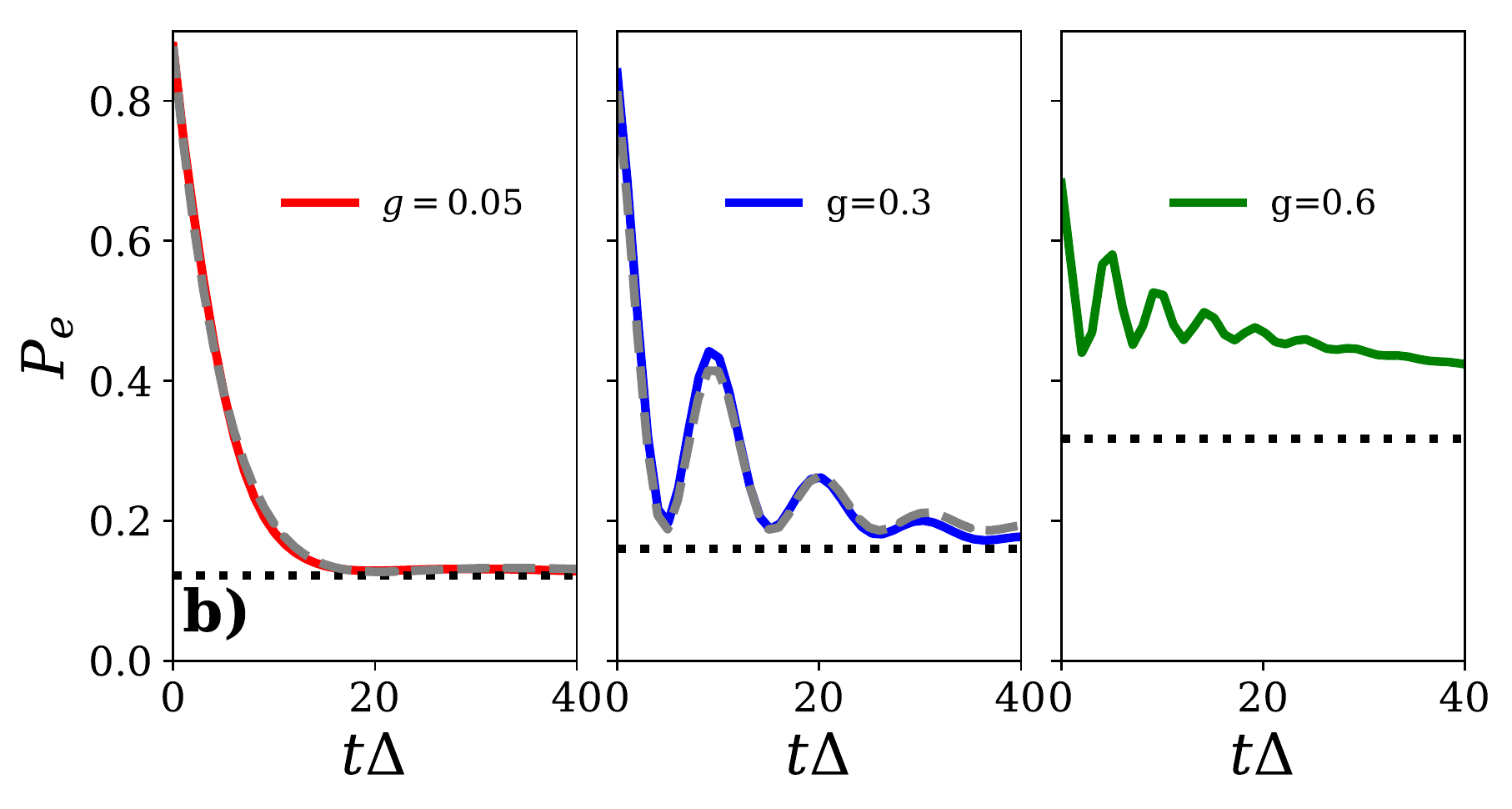}
 \caption{Rabi oscillations. Panel a) is a contour for $P_e(t)$ as a function of the light matter coupling. The parameters used are $\kappa = \pi 0.01 \Omega$ and $\Omega=0.68$. This value for the cavity frequency is chosen for being in resonance with the renormalized qubit frequency, $\Delta_r$ when $g=0.3$ [See inset of Fig. \ref{fig:Dr} a)]. In b) we plot three cuts for $g=0.05, 0.3$ and $0.6$. The bare TLS frequency $\Delta=1$. }
\label{fig:Rabi}
\end{figure*}

In this section we study the dynamics of the cavity-QED setup, solving numerically and analytically the qubit excitation probability $P_e$ \eqref{Pet} with the polaron methods detailed in Sect.\ \ref{sec:wigner-weisskopf} and in App. \ref{app:se}. The first result is the evidence of coherent light-matter (Rabi) oscillations that (i) are resonant around the qubit renormalized frequency $\Delta_r$ and (ii) dampen exponentially with a modified spontaneous emission rate 
\begin{equation}
  \label{gammar}
  \gamma_r \simeq J(\Delta_r)\; ,
\end{equation}
determined by the joint spectral function\ \eqref{gamma} that we introduced in Sect.\ \ref{sec:model}. We find that Rabi oscillations start approximately at the boundary
\begin{equation}
 g \cong |\kappa - \gamma_r|/4.
 \label{critical-g}
\end{equation}
Above this critical value, the TLS and the cavity exchange excitations coherently; below this boundary, the qubit exhibits overdamped exponential decay without oscillations. It is remarkable that this boundary is formally the same as the one in the RWA and Markovian approximation ($g/\Omega, \alpha, \alpha_{\rm cav} \ll 1$ regime) \cite{Auffves2010}, but extends to the USC and WUSC regimes, which do not admit a perturbative treatment.

In Fig.\ \ref{fig:Rabi} we show the qubit dynamics for varying $g$. In \ref{fig:Rabi}a we see the appearance of oscillations for $g \gtrsim |\gamma_r - \kappa| /4 \cong 0.1$. These plots confirm that the TLS dynamical frequency is given by $\Delta_r$; since this resonance changes with $g$, we explore on- and off-resonant oscillations as we increase the coupling strength, for fixed $\Omega$. With the parameters used, $\Omega = \Delta_r =0.68$, which is reached at $g=0.3$. Fig. \ref{fig:Rabi}b shows the qubit dynamics at precised couplings. This includes the resonant case $g=0.3$, which shows resonant-like Rabi oscillations, in the middle plot. For lower coupling $g=0.05$, losses dominate and the TLS dynamics is overdamped ---i.e. exponential decay in the polaron frame---. Since for $\alpha=0.1$, $\Delta_r \neq \Delta$, this is an example of WUSC dynamics. The right-hand plot shows a USC coupling dynamics, with $g=0.6$, where $\Delta_r \cong 0.4$ and the dynamics are non-resonant Rabi-like oscillations.

We have found an analytical approximation that reproduces and explains the TLS-cavity dynamics and the onset of the Rabi oscillations. We take Eq. \eqref{Jsb}, remove the term that is $\mathcal{O}(f^2)$ and modify the rest, replacing the effective displacements $f_k$ with the original couplings $f_k \to c_k$. The solution of these simplified equations is formally identical to the one for $g/\Delta, \alpha, \alpha_{\rm cav} \ll 1$ (Markov and Lorentzian approximations) but with a renormalized frequency $\Delta \to \Delta_r$. We denote it $ \widetilde{P}_e(t)$.
Besides, we need to impose that the time converges in the $t \to \infty$ to the correct equilibrium solution given by $P_e^{\rm eq}= (1 - \Delta_r/\Delta)/2$ [Cf. \eqref{Pet} and \eqref{sz}]. Notice, that our numerical simulations verified thermalization, marked as dotted lines in Fig. \ref{fig:Rabi}. To have the correct stationary limit, we use the simplest interpolation for our analytical estimation,
\begin{equation}
\label{Pea}
P_e^{app} (t) \cong (1- P_e^{\rm eq}) \widetilde{P}_e(t) + P_e^{\rm eq} \; .
\end{equation}
In figure, \ref{fig:Rabi}b) we show how such a approximation holds for relatively high $g$ (well inside the ultrastrong coupling regime), justifying equation \eqref{critical-g}. At strong coupling, $g=0.6$, the simple approximation $P_e^{app}$ stops working (not shown), because we have neglected the $\mathcal{O}(f^2)$ terms. In any case, the evolution still reflects detuned Rabi oscillations, converging to the expected limit: $P_e(t\to \infty) \to \frac{1}{2}(\Delta_r/\Delta - 1)$.

\begin{figure}[b!]
\includegraphics[width=1.\linewidth]{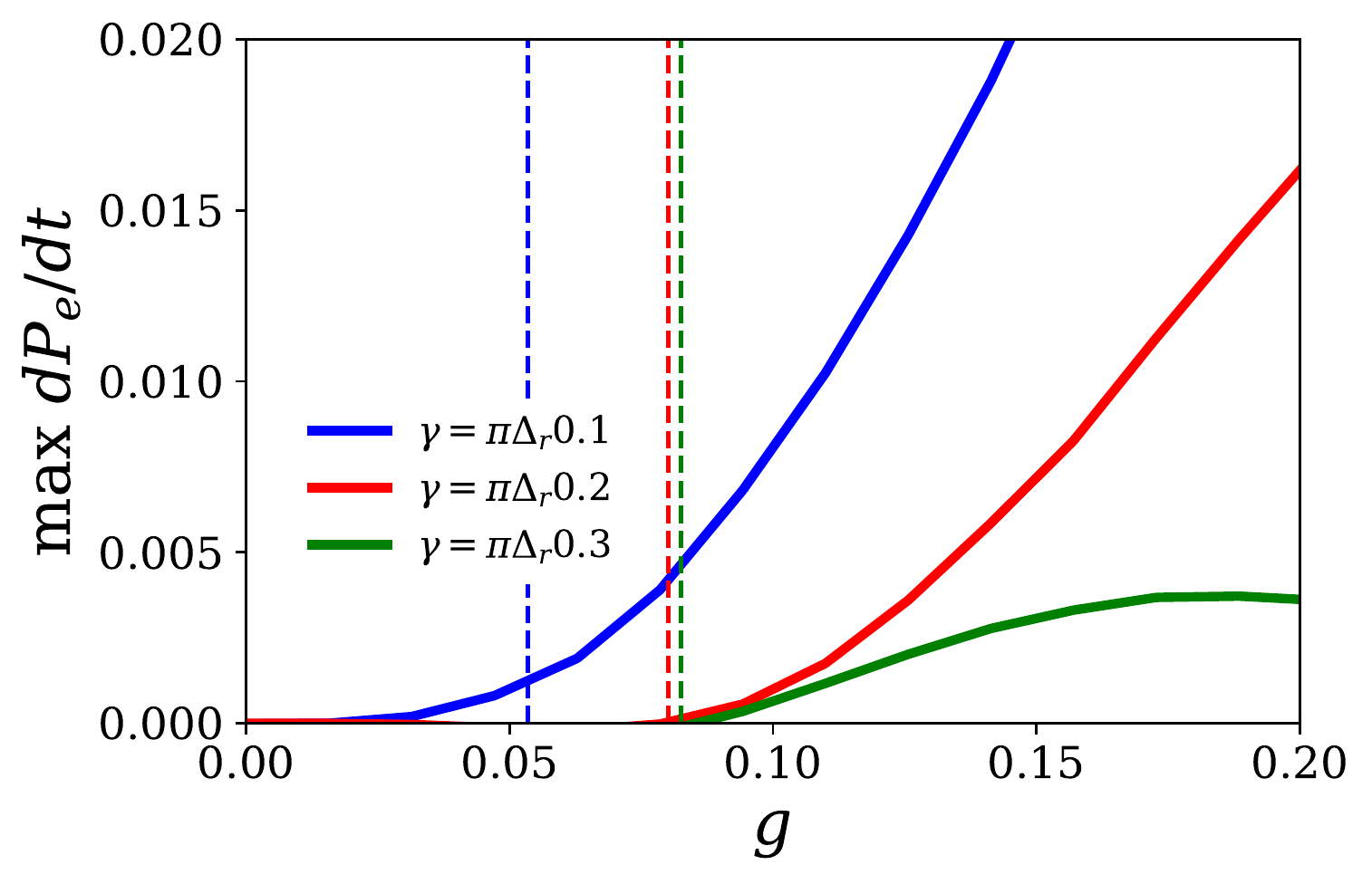}
\caption{Weak-coherent coupling. We plot the maximum derivative of the qubit excitation probability, $\max d P_e / dt$ as a function of $g$ for different noise strenght (solid lines). This derivative is negative or zero in the overerdamped regime. Vertical dashed lines mark the predicition for the critical $g$ given by \eqref{critical-g}. Equal colors mean equal parameters.}
\label{fig:weak}
\end{figure}

We further verified the location of the critical value\ \eqref{critical-g}, analyzing numerically the transition from an overdamped dynamics, to the appearance of the first oscillations. For that we compute the maximum of the time derivative of $P_e$ on the time interval $[0, T]$ with $T$ sufficiently large. We denote this quantity as ${\rm max} \; d P_e / dt$. If the TLS is overdamped, then $d P_e / dt<0$ always and the ${\rm max} \; d P_e / dt=0$. However, if some oscillations occur the derivative is sometimes positive. This is represented in figure \ref{fig:weak} and compared with the bound \eqref{critical-g}, exhibiting a very good agreement. Two comments are in order. First, to generate Fig. \ref{fig:weak} we have chosen to be approximately at resonance at the critical value of $g$. Second, since $\Delta_r$ can go to zero faster that linearly with $\alpha$, $\gamma_r$ approaches to zero by increasing $\alpha$. In the figure, we indeed see that $\alpha_r (\alpha=0.2) \cong \alpha_r (\alpha=0.3)$ [Cf. Fig. \ref{fig:weak}].

Summing up, our simulations justify the use of quantum optics approximations in this non perturbative regime. We find that Rabi oscillations extend qualitatively into a regime where light-matter interaction and dissipation are both non-perturbative. In particular, we have studied a novel regime ---denoted Weak Ultrastrong coupling regime (WUSC)--- in which $g$ is big enough that we cannot make the RWA, but at same time losses are also large and prevent coherent exchange between light and matter degrees of freedom [cf. Fig. \ref{fig:pd}].

\subsection{Qubit noise spectrum}
\label{sec:Sw}

\begin{figure}[b!]
\includegraphics[width=0.8\linewidth]{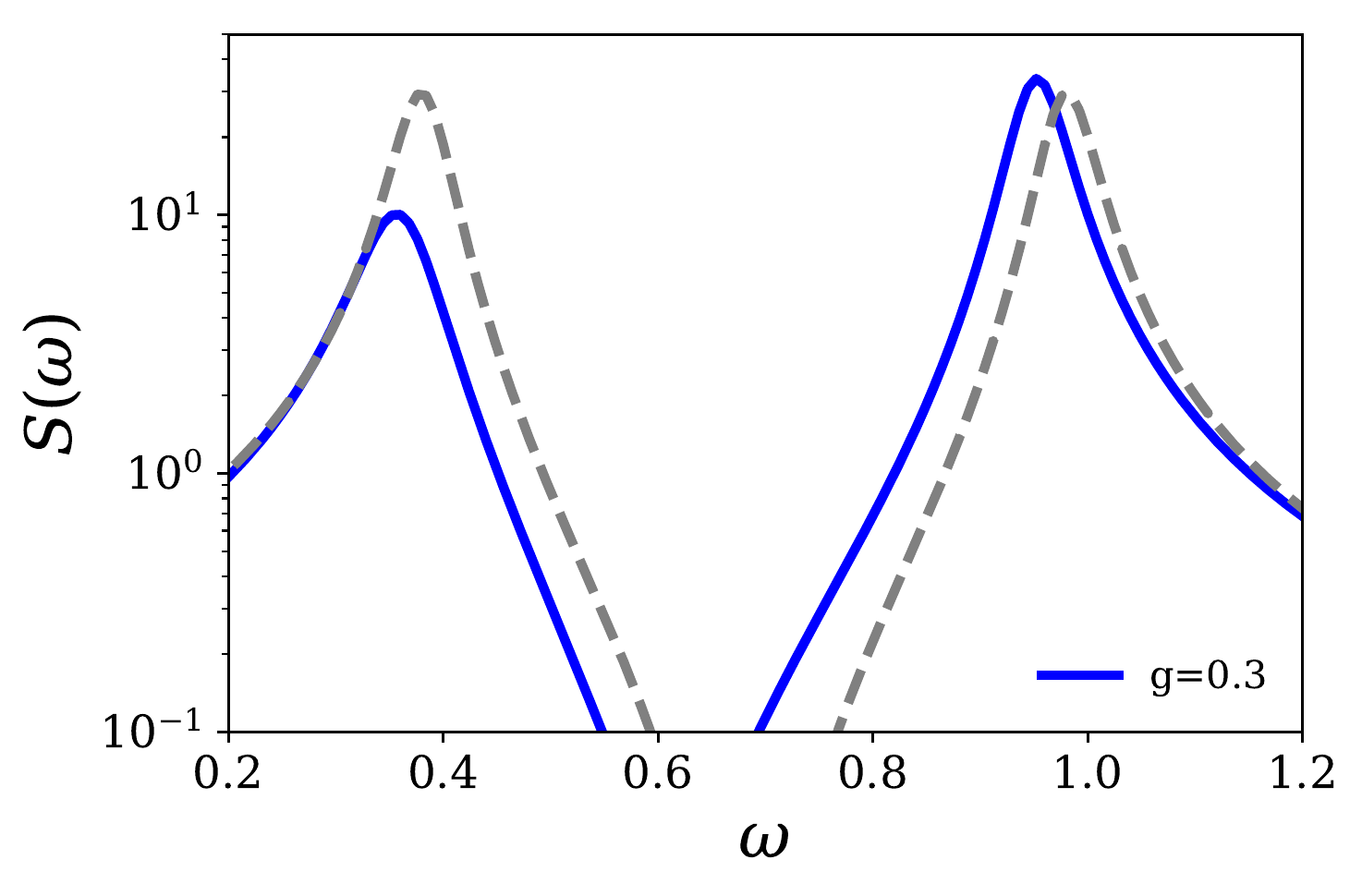}
\caption{$S(\omega)$ at resonance ($\Delta_r=\Omega$). The parameters are those of Fig. \ref{fig:Rabi} $g=0.3$, $\alpha=0.1$, $\Omega = \Delta_r=0.68$ and $\kappa = \pi 0.01 \Omega$.   The blue lines stand for $S(\omega)$ calculated with Eqs. \eqref{Rp} and \eqref{Gp}. The gray dashed lines are calculated with the approximations \eqref{Gmark} and \eqref{Rmark} }
\label{fig:Sw}
\end{figure}
The TLS emission spectrum $S(\omega)$ is a very useful experimental tool that provides information about the coupling $g$ and TLS line-width. $S(\omega)$ is typically computed using the input-output formalism \cite{Gardiner1985}. In this framework, output and input fields are related to the TLS state via $a_{\rm out} = a_{\rm out} - i \sqrt{\Gamma_{\rm TLS}} X^-(t),$ where $\Gamma_{\rm TLS}$ is the emission rate into the transmission line in wich the output signal is collected and $X^-$ is the negative frequency component of the qubit-TL coupling operators ($\sigma_x$ in our case) \cite{Ridolfo2013,Stassi2016}. The emission spectrum is defined as $S(\omega) = \int_0^\infty {\rm d}t \int_0^\infty {\rm d}t^\prime e^{-i \omega (t-t^\prime) }\langle X^+ (t) X^- (t^\prime) \rangle$. Since $H_p$ is number conserving, $S(\omega)$ can be calculated directly from the Laplace transform $\alpha(s)$ in \eqref{WWPa}. After some algebra (fully specified in App. \ref{app:Sw}) we end up with 
\begin{equation}
\label{Sw}
S (\omega) \sim \frac{1}{ \big ( \omega - \Delta_r - R(\omega) \big)^2 + \Gamma(\omega)}
\end{equation}
where $R(\omega)$ and $\Gamma(\omega)$ are the real and imaginary part of the self energy of the qubit.  The former gives the position of the eigenvalues and the latter the line width.  Their explicit expressions ares
\begin{subequations}
\begin{align}
\label{Rp}
R (\omega) = & 
\frac{2 \Delta_r \Big( (\mathcal{K}^\prime)^2 + (\mathcal{K}^{\prime
 \prime})^2 ) - 2 \Delta_r\mathcal{K}^{\prime \prime}\Big )}
{(( 2 \Delta_r)^2 -
 \mathcal{K}^{\prime \prime})^2 + (\mathcal{K}^\prime)^2} 
\\
\label{Gp}
\Gamma (\omega) = & 
\frac{ ( 2 \Delta_r)^2 \mathcal{K}^{\prime }}
{(( 2 \Delta_r)^2 -
 \mathcal{K}^{\prime \prime})^2 + (\mathcal{K}^\prime)^2} 
\end{align}
\end{subequations}
and $\mathcal{K}^{\prime}$ ($\mathcal{K}^{\prime \prime}$) the real (imaginary) part of
\begin{align}
\nonumber
\mathcal{K}
= &
-i
( 2 \Delta_r)^2
 \int_0^\infty \frac{ J (\nu)}{(\nu + \Delta_r)^2} \frac{1}{(\nu
 -\omega) - i 0^+} d \nu \,.
\end{align}
We notice that the renormalized frequency, $\Delta_r$, is again explicit.   If $\alpha, \alpha_{\rm cav}, g/\Delta \ll 1$, the linewidth reduces to
\begin{subequations}
\begin{align}
\Gamma (\omega) &= \frac{g^2 \pi \alpha_{\rm cav} \Omega /2}{(\omega - \Omega)^2 +
 (\pi \alpha_{\rm cav} \Omega  /2)^2} + \pi \alpha \Delta
\label{Gmark}
\\
 R(\omega) &=
\frac{g^2 (\omega-\Omega)}{(\omega - \Omega)^2 +
(\pi \alpha_{\rm cav} \Omega/2)^2} \; ,
\label{Rmark}
\end{align}
\end{subequations}
recovering the standard results in cavity QED (using \emph{e.g.} master equations for dealing with the bath \cite{Carmichael1989}).

The expressions \eqref{Rp} and \eqref{Gp} evidence important corrections in the response profile, with the most evident fact of an asymmetry between peaks at the dressed resonance $\Omega = \Delta_r$ [Cf. Fig. \ref{fig:Sw}]. This is a signature of the modified coupling constants $c_k \to f_k$ in the non perturbative polaron Hamiltonian. From Eq.\ \eqref{fkDr} the effective coupling coefficients $f_k$ are smaller than $c_k$ for $\omega_k < \Delta_r$ and bigger in the region of the spectrum, leading to the asymetric profiles. A similar pheonomenon has been identified in the USC regime, which accounts for large $g$ but only for weak dissipation\ \cite{Cao2011}.

In the previous section we tested a simple approximation to the TLS dynamics\ \eqref{Pet}. This consisted in taking the RWA, weak/strong coupling solutions ---i.e. $\alpha, \alpha_{\rm cav}, g/\Delta \ll 1$---, but replacing the qubit resonance $\Delta \to \Delta_r$, and correcting for the and correct the equilibrium state. If we use this approximation to estimate $S(\omega)$ we find that it more or less accounts for the linewidths. However, our qualitative method fails to reproduce the asymmetry of the peaks [cf. Fig. \eqref{fig:Sw}], and also has a minor error in the peak location, due to the the Bloch-Siegert and dissipation-induced shifts.


\subsection{Circuit QED implementation}
\label{sect:implementation}

The model that we have discussed in this work admits a straightforward realization using superconducting circuits. The three elements that we need are (i) a qubit that is ultrastrongly coupled to the cavity\ \cite{Niemczyk2010,Forn-Diaz2010}, (ii) a possibly ultrastrong coupling between the same qubit and some external environment, such as a transmission line\ \cite{Forn-Diaz2017}, and (iii) a strong or ultrastrong coupling between the superconducting cavity and its own bath, a regime already achieved in\ \cite{haeberlein2015}.

All these three elements admit full and independent tuneability, probing arbitrary values of $g$, $\kappa$ and $\gamma$. First of all, the coupling of the qubit to the photons ---$g$ and $\gamma$ in our model--- can be tuned using SQUIDs that can be embedded in the design of the qubit itself\ \cite{peropadre2010, Peropadre2013}, as demonstrated in Ref.\ \cite{Forn-Diaz2017} for a flux qubit in an open transmission line. Moreover, the coupling between the cavity and its own bath $\kappa$, can also be adjusted using in-line dc SQUIDs. This has been demonstrated in the lab, and used for photon trapping and release\ \cite{yin2013,pierre2014} ---applications that are much more demanding than the simple, stationary tuning of the parameter $\kappa$ in our model.

Assuming a superconducting circuit implementations, we can probe the physics of the combined environments in various ways. We have studied the spectral function $J(\omega)$, which can be reconstructed from the spontaneous decay of an excited qubit. This requires a protocol in which (i) the cavity is decoupled, $\kappa\to0$, (ii) the qubit is excited, (iii) all couplings are switched on for a brief period of time, and (iv) the excited population of the qubit is measured in a non-destructive way\ \cite{reed2010}. Alternatively, it is possible to relate the spectral function to the qubit spectroscopy, by studying the low-power transmission spectrum of the cavity and relating the total lineshape to the spectral function, using the theory from Ref.\ \cite{shi2018}.

\section{Summary}
\label{sec:conclusions}

We have studied a cavity QED model beyond the standard perturbative treatment of losses, using numerical and analytical techniques that apply both in and out of equilibrium. Our study builds on the polaron Hamiltonian\ \cite{shi2018} and Matrix-Product State simulations. In the former case, we have shown that techniques to solve the RWA and weak noise regime (as Wigner-Weisskopf or  $S(\omega)$ calculation) can be extended to work with a variety of regimes ---USC, weak, strong coupling, etc---.

As concrete applications, we have discussed with detail the case of a two-level system that couples ultrastrongly to both the cavity and the bath ---i.e. both $g$ and $\gamma$ are comparable to the qubit and cavity resonances. Using our techniques, we prove that strong dissipation renormalizes the qubit frequency, leading to a new resonance $\Omega=\Delta_r$ and changing its decay rate $\gamma_r$. Our simulations show that this renormalized decay rate can be used to define the onset of Rabi oscillations ($g \cong|\gamma_r - \kappa| /4$), in a formula that extends beyond RWA for all range of parameters. This suggests a new regime where light and matter are ultrastrongly coupled but losses are large enough to suppress Rabi oscillations. We call this regime the \textit{weak ulstrastrong coupling regime} (WUSC) [Cf. Fig. \eqref{fig:pd}].

This work has different possible continuations. On the experimental side, we have shown that all regimes and physics shown in this work can be probed using state-of-the-art circuit QED technology. On the theory side, our numerical methods open the door to extend those experiments to study very challenging cavity QED phenomena, such as transmission-reflection experiments, Dicke physics or nonlinear optics, in the USC and WUSC regimes ---enabling ultrasfast, broadband photon sources, opening access to stronger nonlinearities and facilitating the study of non-Markovian open quantum systems, among other new phenomena to be explored.

\begin{acknowledgments}
We would like to thank Javier Aizpurua, Alejandro Bermudez and Luis Martín-Moreno for inspiring discussions.
We acknowledge support by the Spanish Ministerio de Economia y Competitividad within projects MAT2014- 53432-C5-1-R, FIS2014-55867 and FIS2015-70856-P, and by CAM PRICYT Research Network QUITEMAD+ S2013/ICE-2801.
\end{acknowledgments}


\appendix

\section{ Bath discretization }
\label{app:discrete}

In this section we explain how to discretize the bath for doing the numerical simulations. 

\subsection{The Ohmic case}
\label{sec:ohmic}

In the spin-boson model,
\begin{equation}
\label{sm.Hsb}
H= \frac{\Delta}{2}\sigma^z + \sigma^x \sum_k (g_k a^\dagger_k + \mathrm{H.c.}) + \sum_k \omega_k a^\dagger_k a_k \; ,
\end{equation}
the bosonic bath is fully characterized by the spectral density, defined here as:
\begin{equation}
\label{sm.J}
J(\omega) := 2\pi\sum_k |g_k|^2\delta(\omega-\omega_k) \, .
\end{equation}
In the Ohmic case ($J(\omega) = \pi \alpha \omega$) the spontaneous emission rate in the Markovian regime is given by:
\begin{equation}
\label{sm.G}
\Gamma = J (\Delta) = \pi \alpha \Delta \; .
\end{equation}
The dimensionless parameter $\alpha$ quantifies the spin-boson coupling strenght. 

To find a discretization, \emph{i.e.} a finite set of coupling constants $g_k$, we first rewrite the sum in \eqref{sm.J}:
\begin{equation}
J(\omega) =
2\pi\sum_k \Delta\omega_k \frac{|g_k|^2}{\Delta\omega_k}\delta(\omega-\omega_k)
\end{equation}
so that,
\begin{equation}
\label{sm.Jd}
J (\omega) = 2\pi\frac{|g_k|^2}{\Delta\omega_k} \, ,
\end{equation}
with $\Delta\omega_k$ is the the frequency interval around $\omega_k$. When a frequency is degenerate $\omega_k \simeq \omega_{k'}$, we have to add up all contributions coming from the different couplings.

\begin{figure}
\includegraphics[width=1.\linewidth]{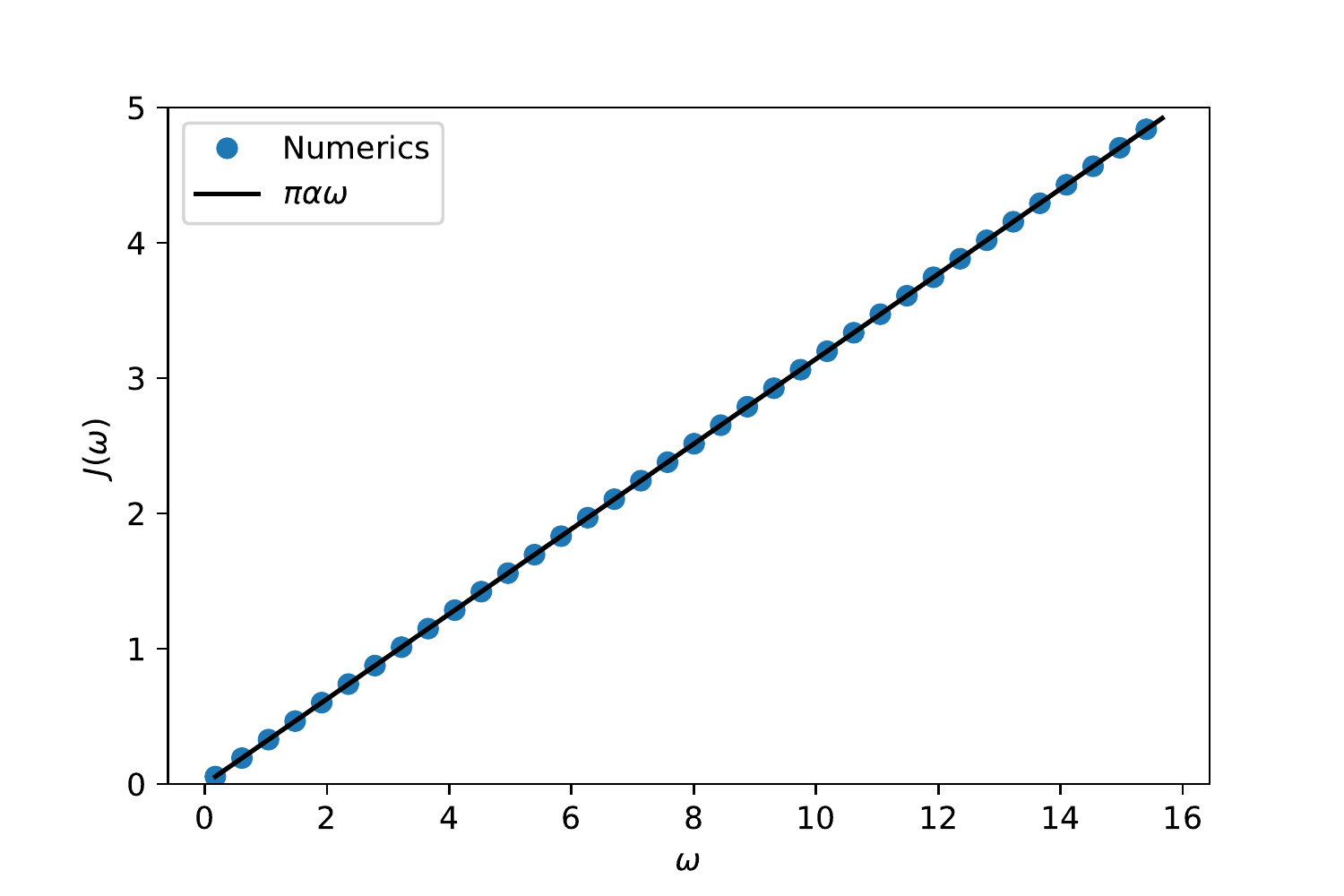}
\caption{Numerical Discrete (circles) versus continumm Ohmic ($J(\omega) = \pi \alpha \omega$) bath. The parameters used are $\alpha=0.1$ and $N=128$ bath modes. For aesthetic reasons, only every 5 points are plotted.}
\label{fig:Jw-Ohmic}
\end{figure}

A transmission line is a model for an Ohmic bath. Its discrete version is a set of coupled harmonic oscillators
\begin{equation}
H_{\rm TL} = \sum \frac{p_i^2}{ 2 \Delta x} + \sum \frac{(x_i - x_{i+1})^2}{2 \Delta x} \; .
\end{equation}
The normal modes are known. In the chiral $k\geq 0$ case,
\begin{equation}
k = \Delta k \times \{0,1,\ldots, N\} \, ,
\end{equation}
where the momentum spacing relates to $\Delta x$
\begin{equation}
\Delta k = \frac{2\pi}{(2N+1)\Delta x}
\end{equation}
which itself dictates the cut-off
\begin{equation}
\Delta x = \frac{2v}{\omega_c}.
\end{equation}
The dispersion relation is then simply obtained from the speed of light ($v=1$)
\begin{equation}
\omega_k = \omega_c \sin(\Delta x k / 2) \, .
\end{equation}
In our model for the TLS $\Delta \omega_k =v \Delta k = \Delta k$, so that using \eqref{sm.Jd} and \eqref{sm.G}
\begin{equation}
g_k = \sqrt{\frac{\alpha \Delta\times \Delta k}{2}}\times \sqrt{\omega_k} = \sqrt{\frac{\Gamma \Delta\times \Delta k}{2\pi}}\times \sqrt{\omega_k}
\; ,
\end{equation}
wich are the coupling used in our numerical simulations.
In figure \ref{fig:Jw-Ohmic} we compare the continuum spectral density and this discretization. The agreement is clear.

\subsection{ The cavity-bath case}
\label{app:cav}

The cavity-bath model is the positive defined quadratic Hamiltonian: 
\begin{equation}
\label{sm.CL}
H = \frac{1}{2}P^2 + \frac{1}{2}\Omega^2 X^2 + \sum_k 
\frac{1}{2}p_k^2 + \frac{1}{2}\omega_k^2 \left( x_k -\frac{c_k}{\omega_k}X \right) ^2
\;.
\end{equation}
This is nothing but the Caldeira-Legget model of dissipation \cite{Caldeira1983}.
The spectral density of the cavity-bath is given by:
\begin{equation}
\label{sm.Jcav}
J_{\rm cav}= \frac{\pi}{2 \Omega} \sum \frac{c_k^2}{\omega_k} \delta (\omega - \omega_k)
\end{equation}
This expression ensures that $\Gamma_{\rm cav}= J(\Omega)$ [Cf. Eq. \eqref{sm.G}]. The differences between this last expression and \eqref{sm.J} arise because \eqref{sm.CL} is written in terms of position-like operators and in the spin boson in terms of annihilation-creation operators.

\begin{figure}
\includegraphics[width=1.\linewidth]{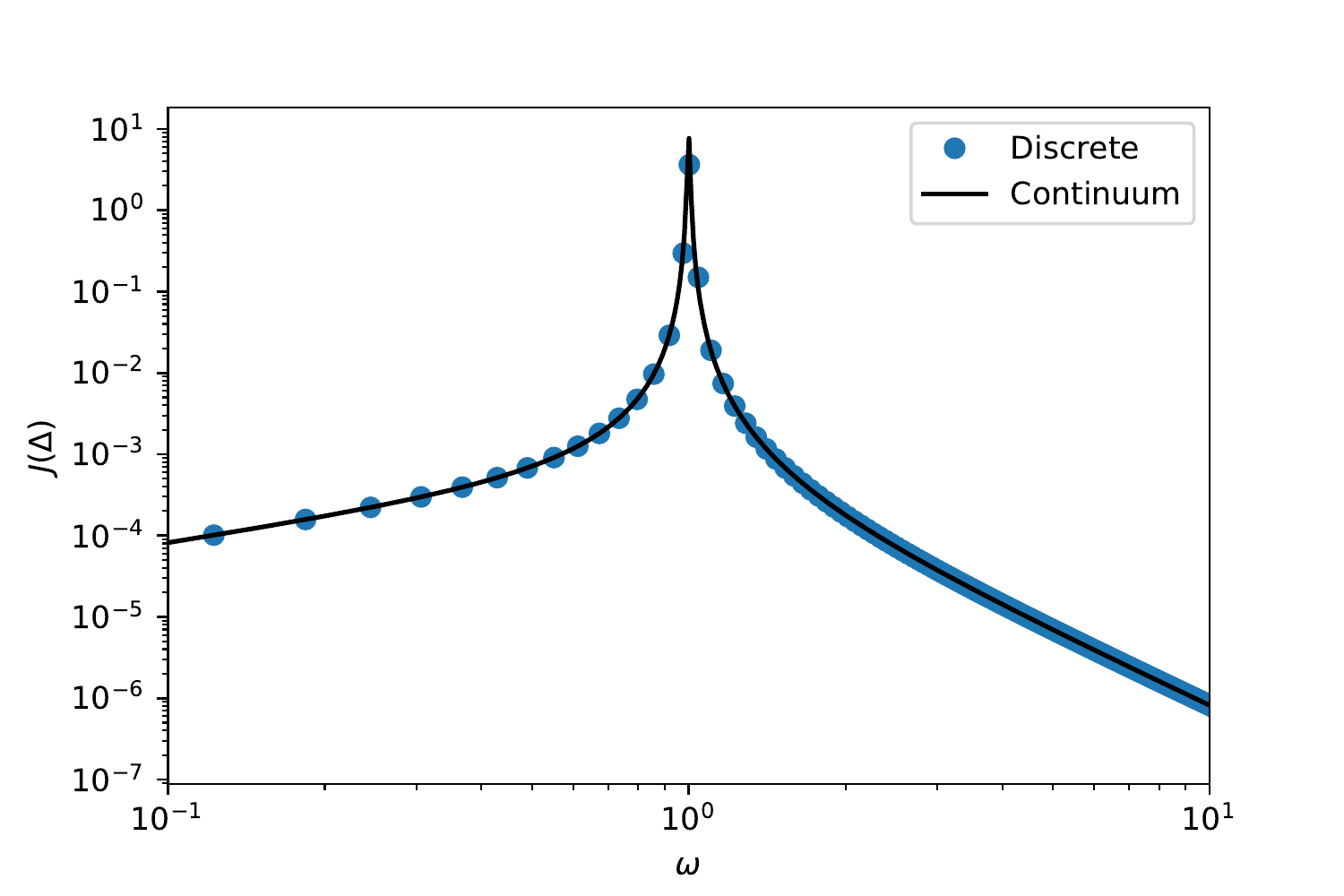}
\caption{Numerical Discrete (circles) versus peaked spectral density continuum bath. The parameters used are $g=0.2$, $\kappa=0.01$ and $N=128$ modes. }
\label{fig:Jw-peaked}
\end{figure}

The Caldeira-Legget model can be rewritten as:
\begin{equation}
\label{sm.Hqfm}
H=\frac{1}{2}(\mathbf{P}^T \mathbf{P} + \mathbf{X}^T B \mathbf{X})
\end{equation}
with the matrix
\begin{equation}
B=\left(\begin{array}{cccc}
\Omega^2 + \sum_k \frac{c_k^2}{\omega_k^2} & -c_1 &\ldots & -c_N\\
-c_1 & \omega_1^2 & 0 & 0\\
\ldots & 0 & \ldots & 0\\
-c_N & 0 & 0 &\omega_N^2
\end{array}\right)
\end{equation}
This model can be further diagonalized using a unitary transformation $U$ and eigenvalues $\Omega^2$ such that
\begin{equation}
\label{sm.U}
B = U \hat\omega^2 U^T
\end{equation}
to give
\begin{equation}
\label{sm.Hqfd}
H = \frac{1}{2}\mathbf{\hat{P}}^T\mathbf{\hat{P}} + \frac{1}{2}\mathbf{\hat{X}}\hat\omega^2\mathbf{\hat{X}} 
\, .
\end{equation}
We quantized the model as usual:
$$\hat{X}_j = \sqrt{\frac{1}{2\hat\omega_j}} (c_i + c_i^\dagger) \; ,$$
obtaining an expansion for the original cavity mode:
\begin{equation}
\label{sm.XUX}
X = (U\mathbf{\hat{X}})_0 = \sqrt{\frac{1}{2}} \sum_{j=1}^{N+1}(U\hat\omega^{-1/2})_{1j}(c_i+c_i^\dagger) \;.
\end{equation}
In doing so, the qubit-cavity coupling can be written as the spin-boson coupling:
\begin{align}
\label{sm.effc}
\nonumber
g \sigma^x (a+a^\dagger) & = g\sqrt{2\Omega}\sigma^x X 
\\
&= g \sigma^x \sum_{j=1}^{N+1}\left(U\sqrt{\frac{\Omega}{\hat\omega}}\right)_{1j}(c_i+c_i^\dagger) \,.
\end{align}
The bath frequencies are given by the eigenvalues $\hat \omega_i$. As we have done with the Ohmic spectral density, in figure \ref{fig:Jw-peaked} we compare the discrete model with the continuum one. The expression for the continuum case is explained in the next section \ref{app:peaked} [See Eq. \eqref{sm.J2}]. Again, the agreement is clear.

\section{ Exact diagonalization and effective spectral density }
\label{app:peaked}

We have explained how to diagonalize the cavity and its bath numerically. It turns out that this task can be also done analytically \cite{Ullersma1966, Ford1988, Ambegaokar2006, Ambegaokar2007}. As a result, the diagonalized cavity mode plus bath can be characterized by a effective spectral density \cite{Garg1985}. We summarize here, in a unified way, both the diagonalization and the effective spectral density.

\subsection{Exact diagonalization}
\label{app:diago}
The
eigenvalue problem \eqref{sm.Hqfm} reads,
\begin{subequations}
\begin{align}
\label{eig-1}
\Omega^2 X - \sum c_k x_k + \sum_k \frac{c_k^2}{\omega_k^2} X & = \hat \omega_j^2 X
\\
\omega_k^2 x_k - c_k X & = \hat \omega_j^2 x_k 
\, .
\label{eig-2}
\end{align}
\end{subequations}
Here, $ \hat \omega_j^2$ are the eigenvalues [Cf. Eq. \eqref{sm.Hqfd}]. From \eqref{eig-2}
\begin{equation}
\label{xkX}
x_k= \frac{c_k}{\hat \omega_j^2-\omega_k^2} \, X \; .
\end{equation}
Inserting the latter in \eqref{eig-1} we obtain:
\begin{equation}
\Omega X + \left ( \sum_k \frac{c_k^2}{\omega_j^2-\omega_k^2 } +
 \frac{c_k^2}{\omega_k^2 } \right )\, X
= \hat \omega_j^2 X \, .
\end{equation}
Therefore, the eigenvalues $\omega_j^2$ are the zeros of the function
\begin{equation}
\label{gm1}
g^{-1} (\omega) = \omega^2 - \Omega^2 - \left ( \sum_k \frac{c_k^2}{\omega_j^2-\omega_k^2 } +
 \frac{c_k^2}{\omega_k^2 } \right )
 \; .
\end{equation}
Defining:
\begin{equation}
\label{Aw}
\mathcal A(\omega) := \left ( \sum_k \frac{c_k^2}{\omega_j^2-\omega_k^2 } +
 \frac{c_k^2}{\omega_k^2 } \right ) \; ,
\end{equation}
we rewrite $g^{-1}$ is a more convenient way, namely:
\begin{equation}
g^{-1} (\omega) = \omega^2 - \Omega^2 - {\mathcal A} (\omega) \; .
\end{equation}
The zeros of $g^{-1}$ are the poles of $g$. It is important to notice that 
 the residues of $g$ are:
\begin{equation}
\label{Res}
{\rm Res} (f, \omega_j) = \frac{1}{\frac{\partial g^{-1}}{ \partial
 \omega} \Big{ |}_{\omega= \omega_j}}
 \; .
\end{equation}

The orthogonal transformation \eqref{sm.U} fulfills the normalization condition:
\begin{equation}
\label{1U}
1 = U_{0 j}^2 + \sum_k U_{kj}^2 = \left ( 1 + \sum_k
 \frac{c_k^2}{(\omega_j^2-\omega_k^2)^2} \right ) U_{0j}^2
 \; .
\end{equation}
In the second equality we have used \eqref{xkX}. Now, we notice that
[Cf. Eq. \eqref{gm1}]
\begin{equation}
\label{dg1}
\frac{\partial g^{-1}(\omega)}{\partial \omega} = 2 \omega \left (
1+ 
\sum_k \frac{C_k^2}{(\omega^2-\omega_k^2)^2} 
\right ) \; .
\end{equation}
Using Eqs. \eqref{Res}, \eqref{1U} and \eqref{dg1} we arrive to:
\begin{equation}
\label{URes}
U_{0j}^2 = \frac{ 2 \hat \omega_j}{\frac{\partial g^{-1}}{ \partial
 \omega} \Big{ |}_{\omega= \hat \omega_j} } = 2 \hat \omega_j \; {\rm Res} (g, \hat \omega_j) \,.
\end{equation}

\subsection{Effective spectral density}
\label{app:Jeff}

We can rewrite the spin-boson coupling \eqref{sm.effc} using \eqref{URes}:
\begin{equation}
\label{sbeff}
g \left(U\sqrt{\frac{\Omega}{\hat\omega}}\right)_{1j}= g \sqrt{\Omega}
\sqrt{ 2 {\rm Res} (g, \omega_j) } \; .
\end{equation}
Therefore, the effective spectral density for the spin-boson reads [Cf. Eq. \eqref{sm.J}]:
\begin{equation}
\label{sm.Jeff}
J_{\rm eff} (\omega)= 2 \pi g^2 \Omega \sum_j 2 {\rm Res} (g, \omega_j) \delta (\omega -\hat \omega_j) \; . 
\end{equation}
Using now \eqref{sbeff} 
for any well behaved function $f(\omega)$, we have that
\begin{align}
\label{sumg}
\nonumber
\int J_{\rm eff} (\omega) f(\omega) &=
4 \pi g^2 \Omega \sum_j {\rm Res} (g, \omega_j) f (\omega_j) 
\\
&=
 4 \pi g^2 \Omega {\rm Im} \left [ \frac{1}{\pi} \int_0 ^{\omega_c} \, d
 \omega 
g (\omega -i 0^+) \right ] \; .
\end{align}
Therefore:
\begin{equation}
\label{Jeffg}
J_{\rm eff}
(\omega ) = 4 g^2 \Omega {\rm Im} [
g (\omega -i 0^+) ] \; .
\end{equation}

So far, everything was general (we have not specified the spectral
density for the cavity bath). We particularize to an Ohmic
spectral density, see \eqref{sm.Jcav} and \eqref{kappa}: 
\begin{equation}
J_{\rm cav} (\omega) = \pi \alpha_{\rm cav} \omega 
\to \kappa = J_{\rm cav} (\Omega)
\end{equation} \; .
Now, 
we can compute $\mathcal A$ defined in \eqref{Aw}
\begin{align}
\mathcal A(\omega) &=
\frac{2}{\pi} \Omega \int d \nu
\frac{ J (\nu) \nu}{\omega^2 - \nu^2} - \frac{J (\nu)}{\nu} 
\\ \nonumber
&= 
 \frac{2}{\pi} \Omega \alpha_{\rm cav} \int d \nu \frac{\omega^2 }{\omega^2 - \nu^2} 
 \\ \nonumber
&= i \pi \Omega \alpha_{\rm cav} \omega = i \kappa \omega
\end{align}
Inserting the last result in the definition of $g^{-1}$, taking the
imaginary part and using \eqref{Jeffg} we get 
\begin{equation}
\label{sm.J2}
J_{\rm eff} (\omega) = 
\frac{ 4 g^2 \kappa
 \omega}{(\Omega^2-\omega^2)^2 + (\kappa \omega)^2} 
 \; .
\end{equation}
which is nothing but the peaked spectral density discussed in the main text and that has been used to test our bath discretization [Cf. Fig. \eqref{fig:Jw-peaked}].


\section{Single excitation time evolution (analytical calculations) }
\label{app:se}

We approximate Eqs. \eqref{WWPa} and \eqref{WWPb} as follows.
We remove the term that is $\mathcal{O}(f^2)$ and modify the rest, replacing the effective displacements $f_k$ with the original couplings $f_k \to c_k$.  Besides, working in the rotated basis
$ \widetilde \psi = e^{i \Delta_r t} \psi $ and $ \widetilde \psi_k = e^{i \Delta_r t} \psi_k $ ) these dynamical equations yield: 
\begin{subequations}
\begin{align}
\label{sm.WWPa}
\dot {\widetilde \psi} &= - i 2 \Delta_r \sum \widetilde \psi_k f_k
\\
\label{sm.WWPb}
\dot {\widetilde \psi_k} &= -i (\omega_k - \Delta_r) \widetilde \psi_k - i 2 \Delta_r c_k  \widetilde \psi 
\; ,
\end{align}
\end{subequations}
The set of amplitudes $\widetilde \psi_k$ can be formally integrated and replace the solutions in the equation for $\widetilde \psi$ arriving to the non-local differential equation:
\begin{equation}
 \label{sm.dta}
\dot {\widetilde \psi} = - \frac{1}{2\pi}\int^\infty_0 d \omega \int^t_0 d \tau J(\omega) e^{i (\Delta-\omega)(t-\tau)} \widetilde \psi (\tau) \; .
\end{equation}
We recall that $J(\omega)$ is the sum of two contributions, the one coming from the intrinsic TLS noise and the second coming from the cavity [Cf. Eqs. \eqref{Jsb}].
In the markovian limit (wich is consistent with the regime we are discussing), the Ohmic intrinsic TLS dissipation produces a local-term:
\begin{equation}
 - \frac{1}{2 \pi}\int^\infty_0 d \omega \int^t_0 d \pi \psi \omega e^{i (\Delta-\omega)(t-\tau)} \widetilde \psi (\tau)
 \cong - \frac{\pi}{2} \alpha \Delta \widetilde \psi \, .
 \end{equation}
The second summand is approximated with a Lorentzian. Using that $\kappa = \pi \alpha_{\rm cav} \Omega$ [Cfs. Eq. \eqref{Jsb} and \eqref{kappa}]:
 \begin{equation}
 \label{sm.lor}
 \frac{ 4 g^2 \kappa  \Omega \omega }{(\Omega^2-\omega^2)^2 + (\kappa \omega)^2}
 \cong  \frac{g^2 \kappa \Omega}{(\Omega -\omega)^2 + \kappa^2/4}
 \end{equation}
 Because, it is peaked around $\Omega$, we can extend the frequency integral: $\int_0^\infty d\omega \to \int_{-\infty}^\infty d \omega $. We use the Fourier transform of the Loretzian and back to the nonrotated picture ending up with:
 \begin{widetext}
 \begin{equation}
 \dot {\psi}
 = - g^2 \int d t
 e^{i (\Delta_r-\Omega) (t-\tau)} e^{-\kappa |t -\tau|/2}
  \psi (\tau) - \gamma \psi/2 \, .
 \end{equation}
 Taking the time derivative, we have the \emph {local} second order differential equation:
 \begin{equation}
  \ddot {\psi} = -(g^2 - \gamma \kappa/4 - i \delta \gamma/2) \psi - (\gamma + \kappa + 2 i \delta)/2 \dot \psi \; ,
 \end{equation}
 with $\delta = \Delta_r - \Omega$.
 The solution is ($\psi(0)=1$ and $\dot \psi(0) = - \gamma/2$):
 \begin{equation}
  \psi=
  \frac{e^{-\frac{1}{4} t \left(\kappa+\gamma+\eta \right)}
 \left((\gamma- \kappa)(1- e^{\frac{1}{2} \eta t})+ \eta + \eta e^{\frac{1}{2}
 t \eta}\right)}{2
 \eta} \, ,
\end{equation}
where
\begin{equation}
 \eta := \sqrt{(\gamma - \kappa)^2-16 g^2}\; .
 \end{equation}

\section{ Qubit emission spectrum }
\label{app:Sw}

In computing the noise spectrum $S(\omega)$ several manipulations can be made. It is convenient to 
solve  \eqref{WWPa} and \eqref{WWPb} using the Laplace transform.  We do it in the rotated frame $ \widetilde \psi = e^{i \Delta_r t} \psi $ and $ \widetilde \psi_k = e^{i \Delta_r t} \psi_k $. In the $s$-domain the dynamical equations read:
\begin{subequations}
\begin{align}
s \widetilde \psi (s) - \widetilde \psi (0) & = - i 2 \Delta_r \sum_k \psi_k(s) f_k
\\
s \widetilde \psi_k (s) - \widetilde \psi_k (0) &= - i (\omega_k - \Delta) \widetilde \psi_k(s) - i 2 \Delta f_k \Big ( \widetilde \psi(s) + \sum_{k^\prime} f_{k^\prime} \widetilde \psi_{k^\prime} (s)\Big )
\end{align}
\end{subequations}
Using that the qubit is initially excited, $ \psi(0)=1$, and the bath is its ground state, $\psi_k(0)=0$, we find:
\begin{equation}
\widetilde \psi(s) 
=
\frac{1}{s + (2 \Delta)^2 \frac{\sum_k f_k^2/(s+ i (\omega_k - \Delta))}{1+ i 2 \Delta \sum_k f_k^2/(s+ i (\omega_k - \Delta))}} \; ,
\end{equation}
\end{widetext}
which can be written as 
\begin{equation}
\label{bsp}
\widetilde \psi (s) = 
\frac{ 1 }
{s + K (s) 
} \; ,
\end{equation}
where
\begin{equation}
\label{Kp}
K(s)=
 ( 2 \Delta_r)^2
\frac{ \mathcal{K} (s)}{1- i \mathcal{K} (s)/ 2 \Delta_r }
\end{equation}
and
\begin{equation}
\label{sm.K}
\mathcal{K} (s)
= \sum \frac{ (2 \Delta_r)^2 f_k^2}{s + i ( \omega_k - \Delta_r )}
\; .
\end{equation}

The $s$-domain is specially useful for computing the emission espectrum, that in our case is given by:
\begin{equation}
\label{Sw}
S (\omega) 
= \int_0^\infty {\rm d}t \int_0^\infty {\rm d}t^\prime e^{-i \omega
 (t-t^\prime) }\langle \sigma^+ (t) \sigma^- (t^\prime) \rangle \; .
\end{equation}
In the single excitation subspace we have that
\begin{equation}
\label{regr}
\langle \sigma^+ (t +\tau) \sigma^- (t) \rangle 
= \psi^*(t+\tau) \psi(t) \, .
\end{equation}
Using the inversion formula 
\begin{equation}
f(t) = \frac{1}{2 \pi} \int^{\infty}_{-\infty} \; {\rm d} \omega e^{i
 \omega t} f (i \omega + 0^+) \; ,
\end{equation}
we get [Cf. Eq. \eqref{bsp}]
\begin{align}
\widetilde \psi(t) 
= 
\frac{1}{2 \pi}
\int^{\infty}_{-\infty} \; {\rm d} \omega 
\frac{e^{i
 \omega t}}{i \omega + 0^+ + K(i \omega + 0^+)} \, .
\end{align}
Since, $\widetilde \psi
= e^{i \Delta t} \psi$, the relevant object is:
\begin{equation}
\label{beta-int}
\psi (t) = 
\frac{1}{2 \pi i }
\int^{\infty}_{-\infty} \; 
{\rm d} \omega 
\frac{
e^{-i
 \omega t}}
{
\Delta - \omega -i K(i (\Delta_r-\omega) + 0^+)} \, .
\end{equation}
If we split in real and imaginary parts the Kernel:
\begin{equation}
\label{KRG}
K (i (\Delta - \omega) + 0^+) = i [R (\omega) - i \Gamma(\omega)]
\, ,
\end{equation} 
together with \eqref{regr} and the definition
\eqref{Sw}, we realize that:
\begin{equation}
\label{Sw-f}
S (\omega) \sim \frac{1}{ \big ( \omega - \Delta - R(\omega) \big)^2 + \Gamma(\omega)}
\, .
\end{equation} Thus, we \emph{just} to give explicit results for $G(\omega)$ and $\Gamma(\omega)$.

We use Eqs. \eqref{Kp} and \eqref{sm.K}:
\begin{widetext}
\begin{align}
\nonumber
\mathcal{K}(i (\Delta_r - \omega) +
0^+)
= &
-i
( 2 \Delta_r)^2
 \int_0^\infty \frac{ J (\nu)}{(\nu + \Delta_r)^2} \frac{1}{(\nu
 -\omega) - i 0^+} d \nu
\\ 
= &
\pi \frac{ ( 2 \Delta_r)^2 J (\omega)}{(\omega + \Delta_r)^2}
- i ( 2 \Delta_r)^2 \; \mathcal {P} \int_0^\infty \frac{ J (\nu)}{(\nu + \Delta_r)^2} \frac{1}{(\nu
 -\omega) } d \nu
\equiv
\mathcal{K}^\prime -i \mathcal{K}^{\prime \prime} \, .
\label{mathKri}
\end{align}
With Eqs. \eqref{Kp} and \eqref{mathKri}
we have that [Cf. Eq. \eqref{KRG}]:
\begin{align}
\label{Rp}
R (\omega) = & 
\frac{2 \Delta_r \Big( (\mathcal{K}^\prime)^2 + (\mathcal{K}^{\prime
 \prime})^2 ) - 2 \Delta_r\mathcal{K}^{\prime \prime}\Big )}
{(( 2 \Delta_r)^2 -
 \mathcal{K}^{\prime \prime})^2 + (\mathcal{K}^\prime)^2} \, ,
\\
\label{Gp}
\Gamma (\omega) = & 
\frac{ ( 2 \Delta_r)^2 \mathcal{K}^{\prime }}
{(( 2 \Delta_r)^2 -
 \mathcal{K}^{\prime \prime})^2 + (\mathcal{K}^\prime)^2} \, .
\end{align}

\subsubsection{Calculations in the good cavity limit}

To produce analytical results we must solve the principal part in \eqref{mathKri}. In the limit of good cavity $ \alpha_{\rm cav} \ll 1$ the
integral can be done analtytically. Notice that $\Delta_r$ acts as an effective
cutoff. The integral with the peaked part of $J(\omega)$ can not be done
in general. However, if $\kappa$ is small enough $J_{\rm eff}$ is approximated by a Lorentzian \eqref{sm.lor}.
Besides, we can approximate $\nu + \Delta_r \to \Omega + \Delta_r$ in
the denominator. Putting all together 
we have a close formula for $\mathcal K$
\begin{align}
\label{mKgc}
\mathcal{K}(i (\Delta_r - \omega) + 0^+)
=
i 4 \Delta_r^2
 \left [ 
\frac{ R_2^{(0)} (\omega)}{(\Omega+ \Delta_r)^2}
+
\frac{\gamma \big ( \Delta_r + \omega + \omega \log (
 \Delta_r/\omega) \big ) } {(\Delta_r + \omega)^2}
\right .
\left .
- 
i \left (
\frac{ \Gamma_2^{(0)} (\omega)}{(\Omega+ \Delta_r)^2} 
+
\frac{ \pi \gamma \omega }{(\Delta_r + \omega)^2}
\right )
\right ]
\end{align}
where,
\begin{align}
\label{R2w}
R^{(0)}_2(\omega) & = \frac{g^2 (\omega-\Omega)}{(\omega - \Omega)^2 +
    (\kappa/2)^2} \, ,
\\
\label{G2w}
\Gamma^{(0)}_2 (\omega) & = \frac{g^2 \kappa/2}{(\omega - \Omega)^2 +
    (\kappa/2)^2} \, .
\end{align}
\end{widetext}


\bibliographystyle{apsrev4-1}
\bibliography{dissqR}

\end{document}